\documentclass{article}[12pt,psfig]
\usepackage{amsmath}
\usepackage{latexsym}
\usepackage{float}
\usepackage{amssymb}
\usepackage{graphicx}
\usepackage{epsfig}
\usepackage{cite}
\begin{document}
\title{Amorphous silica at surfaces and interfaces: simulation studies}
\author{%
 J. Horbach$^{\text{1)}},\;$
 T. St\"uhn$^{\text{1)}},\;$
 C. Mischler$^{\text{1)}},\;$\,%
 W. Kob$^{\text{2)}}$, and
 K. Binder$^{\text{1)}}$ \\[\baselineskip]%
                   $^{\text{1)}}$\textit{ Institut f\"{u}r Physik, Johannes Gutenberg-Universit\"{a}t,} \\
                   \textit {D-55099 Mainz, Staudinger Weg 7, Germany}\\%
                   $^{\text{2)}}$\textit{ Laboratoire des Verres, Universit\'e
                                   Montpellier II,}\\
                                  \textit{F-34095 Montpellier,
                                  France}\\
\date{}
}
 \maketitle

\begin{abstract}
The structure of surfaces and interfaces of silica (SiO$_2$) is
investigated by large scale molecular dynamics computer simulations. In
the case of a free silica surface, the results of a classical molecular
dynamics simulation are compared to those of an {\it ab initio} method,
the Car--Parrinello molecular dynamics. This comparative study
allows to check the accuracy of the model potential that underlies the
classical simulation. By means of a pure classical MD, the interface
between amorphous and crystalline SiO$_2$ is investigated, and as a
third example the structure of a silica melt between walls is studied in
equilibrium and under shear. We show that in the latter three examples
important structural information such as ring size distributions can be gained from 
the computer simulation that is not accessible in experiments.
\end{abstract}
\section{Introduction}

The understanding of the structural and dynamic properties of silica at
surfaces and interfaces is of special importance for technologically
interesting systems such as semiconductor devices and nanoporous
materials~\cite{legrand98,iler79,helms93}.  Furthermore, since amorphous
silica surfaces are very reactive, they are used in catalysis and
chromatography. Although there are different experimental methods
such as infrared and Raman spectroscopy, atomic force microscopy and
NMR to study surface properties, at least in the case of silica it is
still difficult to make conclusive statements about the microscopic
surface structure of SiO$_2$ from experiments that use the latter
methods~\cite{grabbe95,gupta00,poggemann01,chuang96}.

In order to shed more light on these issues, a molecular dynamics (MD)
computer simulation is a very useful tool. In this method Newton's
equations of motion are solved and thus one obtains the trajectories
of the particles in the system from which one can calculate all the
relevant quantities to characterize the microscopic structure (and
dynamics). However, the predictive power of a MD simulation depends
strongly on the quality of the potential with which one models the
interactions between the particles. One possibility to check the
accuracy of the model potential is a comparison to experiment. But,
especially in the case of the surface structure of amorphous silica,
there is a lack of experimental data. An alternative method to check
the potential is presented in the next section: We use a combination of
classical MD with a model potential and an {\it ab initio} simulation
technique, the so--called Car--Parrinello MD.  The use of classical MD
for the investigation of the non--bulk behavior of amorphous SiO$_2$
is then presented for two other subjects, namely the structure of an
interface between amorphous and crystalline SiO$_2$ (Sec.~3) and the
behavior of amorphous silica between two walls in equilibrium and under
shear (Sec.~4).  Finally we summarize the results in Sec.~5.
\vspace*{-0.2cm}

\section{Free silica surfaces: {\it Ab initio} and classical molecular dynamics}

As a model to describe the interactions between the atoms in silica
we use a pair potential which has been proposed by van Beest {\it
et al.}~\cite{beest90}.  It contains a Coulomb part and a
short--ranged part,
\begin{equation}
  \phi(r)=
    \frac{q_{\alpha} q_{\beta} e^2}{r} +
     A_{\alpha \beta} \exp\left(-B_{\alpha \beta}r\right) -
    \frac{C_{\alpha \beta}}{r^6}\quad \alpha, \beta \in
    [{\rm Si}, {\rm O}] \ ,
    \label{eq1}
\end{equation}
where $r$ is the distance between an atom of type $\alpha$ and an
atom of type $\beta$. The effective charges are $q_{{\rm O}} = -1.2$
and $q_{{ \rm Si}} = 2.4$, and the parameters $A_{\alpha \beta}$,
$B_{\alpha \beta}$, and $C_{\alpha \beta}$ can be found in the original
publication. They were fixed by using a mixture of {\it ab initio}
calculations and classical lattice dynamics simulations. The long--ranged
Coulomb forces (and the potential) were evaluated by means of the Ewald
summation technique. As an integrator for the simulation we used a
velocity Verlet algorithm~\cite{binder03} with a time step of $1.6$~fs.
More details on the potential and the calculation of the forces can be
found in Ref.~\cite{mischler}.

Although the BKS potential has turned out to be very good in
the description of {\it bulk properties} of amorphous silica (see
Ref.~\cite{horbach99} and references therein), it is much less obvious
that this potential is also reliable to model silica surfaces. The
reason is that it was optimized to reproduce bulk properties such as the
experimental elastic constants of $\alpha$ quartz. A similar fitting of
the potential parameters to surface properties of real silica is difficult
because of the lack of experimental data in this case. Although it is
straightforward to investigate the properties of free surfaces of SiO$_2$
by means of a MD using the BKS potential, it is not clear how one could
test whether the results have anything in common with real silica.

One possibility to circumvent this problem is to use Car--Parrinello
molecular dynamics (CPMD)~\cite{car85} in which, different from a
classical MD, the electronic degrees of freedom are taken into account via
a density functional theory. Therefore, in contrast to a classical MD an
effective potential between the ions is calculated self--consistently
on the fly, i.e.~the instantaneous geometry of the ions is always
taken into account and one does not have to model the potential energy
between the ions by a given function such as Eq.~(\ref{eq1}). However,
it is not yet possible to replace the classical MD method by the CPMD,
since due to the huge computational burden only relatively short time
scales, a few ps, as well as small systems, a few hundred particles,
can be simulated. In contrast to that classical MD allows one to simulate
thousands of particles over several ns~\cite{binder03}.

Since the time scale which is accessible in a CPMD is very restricted the
idea of our approach is to combine a classical MD and a CPMD (see also
Ref.~\cite{benoit00}). To this end, we first equilibrate the system with a
classical MD in which we again model the interactions between the atoms by
the BKS potential, Eq.~(\ref{eq1}). Then, we use these configurations as
the starting point of a CPMD. The goal of this investigation is twofold:
Firstly, we want to see whether the classical configurations are stable
in CPMD. If this is the case we can subsequently compare the structural
differences as obtained by the two methods. So we get an idea of how
accurate the BKS potential is able to describe silica surfaces, and we
have hints of how this potential energy model could be improved.

\begin{figure}[t]
\psfig{file=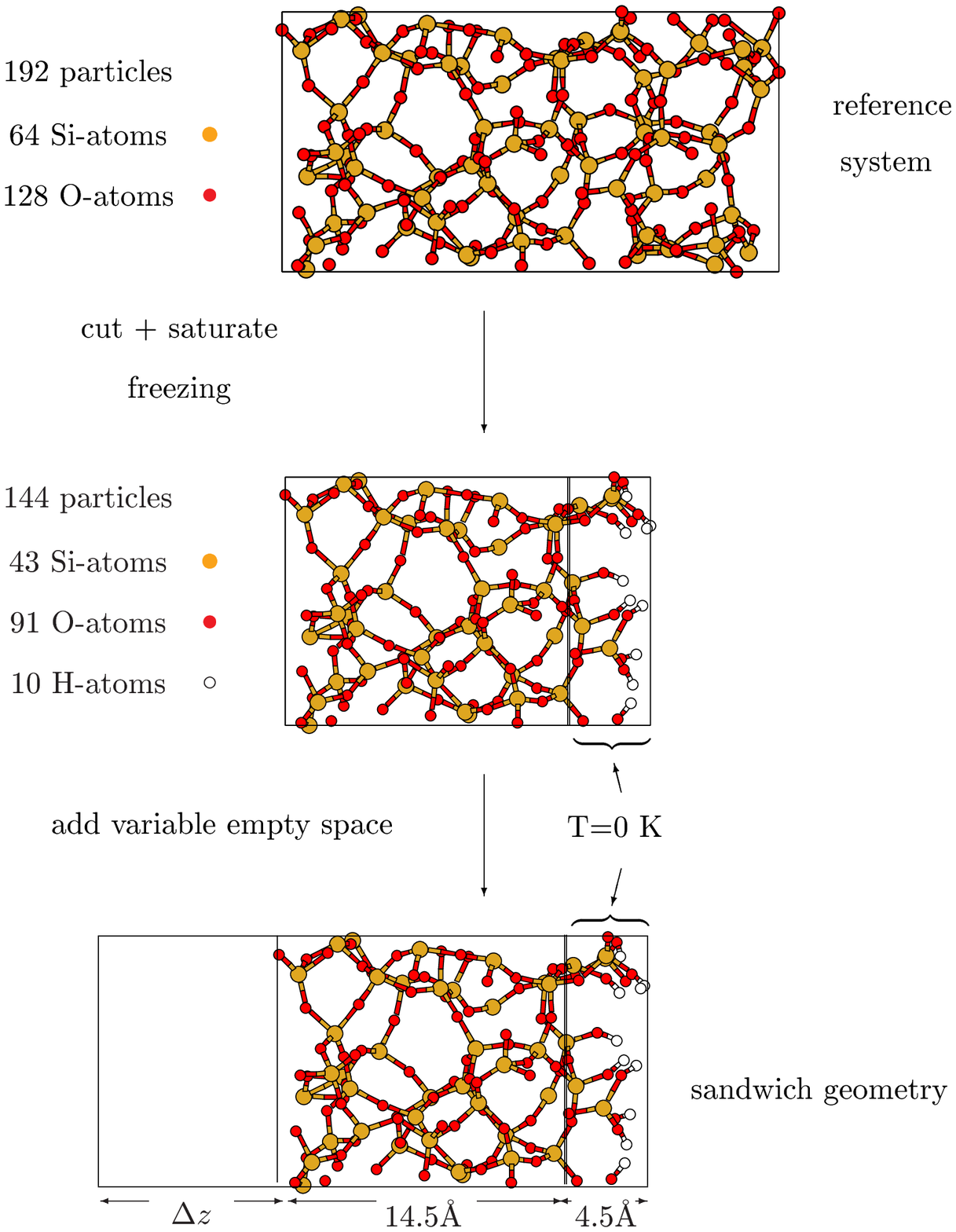,height=18cm}
\vspace{-1cm}
\caption{\label{fig5}Procedure to create the used sandwich geometry.}
\end{figure}

In order to investigate a free silica surface one could consider
a film geometry, i.e.~periodic boundary conditions (PBC) in two
directions and an infinite free space above and below the remaining
third direction. Unfortunately, this is not a very good solution since
the Ewald summation technique for the long ranged Coulomb interactions
becomes inefficient in this case~\cite{arnold02}. Therefore, we have
adopted the following strategy which is illustrated in Fig.~\ref{fig5}:
i) We start with a system at $T=3400$~K with PBC in three dimensions (box
dimensions: $L_x=L_y=11.51$~\AA~and $L_z^{\prime}=23$~\AA). ii) We cut the
system perpendicular to the $z$--direction into two pieces. This cut
is done in such a way that we get only free oxygen atoms at
this interface. iii) These free oxygen atoms are now saturated by hydrogen
atoms. The place of these hydrogen atoms is chosen such that each of the
new oxygen--hydrogen bonds is in the same direction as the oxygen--silicon
bonds which were cut and have a length of approximately $1$~\AA. The
interaction between the hydrogen and the oxygen atoms as well as the silicon
atoms are described only by a Coulombic term. The value of the effective
charge of the hydrogen atoms is set to 0.6 which ensures that the system
is still (charge) neutral. iv) Atoms for which the distance from
the interface that is less than $4.5$~\AA~are made immobile whereas atoms
that have a larger distance can propagate subject to the force field. v)
We add in $z$--direction an empty space $\Delta z = 6.0$~\AA~and thus
generate a free surface at around $14.5$~\AA. With this sandwich geometry
we now can use periodic boundary conditions in all three directions.  We have
made sure that the value of $\Delta z$ is sufficiently large that the
results do not depend on it anymore~\cite{mischler}. Eventually, we have a system of
91 oxygen, 43 silicon, and 10 hydrogen atoms in a simulation box with
$L_x=L_y=11.51$~\AA~and $L_z\approx 25$~\AA.

We have fully equilibrated 100 independent configurations for about
1~ns. Using a subset of these configurations as starting points we
subsequently started CPMD simulations~\cite{cpmd}. We used conventional
pseudopotentials for silicon and oxygen and the BLYP exchange
functions~\cite{trouiller,lee}. The electronic wave--functions were
expanded in a plane wave basis set with an energy cutoff of 60~Ry and
the equations of motion were integrated with a time step of 0.085~fs for
0.2~ps. More details on the CPMD simulations can be found in Ref.~\cite{mischler}.

In the analysis of the CPMD run only those configurations were
taken into account that were produced later than 5~fs after the start
of the CPMD run in order to allow the system to equilibrate at least
locally~\cite{benoit00}. From the 100 BKS configurations we have picked up
those for which one of the three following cases holds: i) The surface
exhibits no defects, i.e.~all Si and O atoms are four and two--fold
coordinated, respectively. ii) The surface has an undercoordinated
oxygen atom and an undercoordinated silicon atom.  iii) The surface has
an overcoordinated oxygen atom and an undercoordinated silicon atom. We
used two BKS configurations for each of the cases i)--iii) and started
the CPMD runs. The computational effort for the latter was very large: To
simulate 1~ps one needs 11000~CPU hours of single processor time (or
about one month on 16 processors of a Cray T3E).

\begin{figure}[t]
\psfig{file=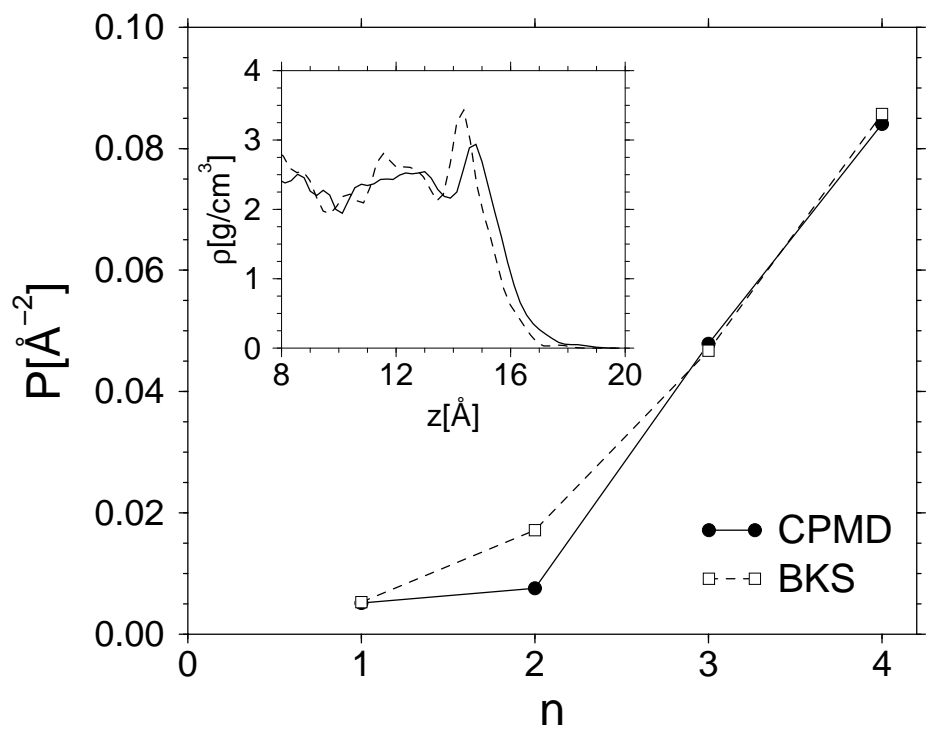,height=10cm}
\caption{\label{fig6}Probability to find a ring of size $n$. 
         Inset: $z$--dependence of the mass density.}
\end{figure}

A quantity which is appropriate to characterize the network structure
is the distribution of the ring size. A ring is defined as a closed loop of $n$
consecutive Si--O segments. The largest differences between the results
of the classical and that of the CPMD simulation is found for the short
rings, i.e.~$n<5$.  Fig.~\ref{fig6} shows the probability to find a
ring of size $n$ for the case of the BKS simulation and the CPMD. We
see that with the BKS model the frequency with which a ring of size
two occurs is overestimated by about a factor of two as compared to
the CPMD result. In agreement with this observation we find that the
overshoot in the $z$--dependence of the mass density profile near the
surface is less pronounced for the CPMD than for the classical MD with
the BKS potential (see inset of Fig.~\ref{fig6}), since two--membered
rings are relatively dense.

\begin{figure}[t]
\psfig{file=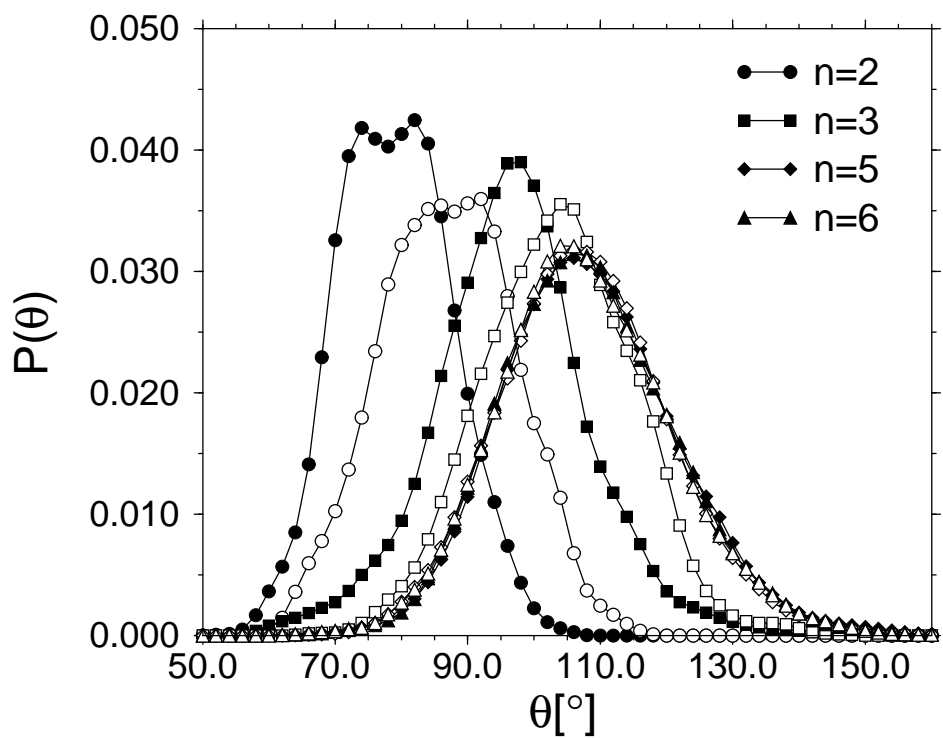,height=10cm}
\caption{\label{fig7}Distribution of O--Si--O angles for different ring sizes (BKS: 
         filled symbols, CPMD: open symbols).}
\end{figure}

Another interesting result is the dependence of the distribution of
angles O--Si--O on the ring size $n$ (Fig.~\ref{fig7}). For large $n$,
$n>4$, i.e.~for rings which are normally found in the bulk, the results
of the two different methods are in good agreement~\cite{benoit00}. For
smaller $n$, however, the mean O--Si--O angle from CPMD is shifted to
larger values in comparison to that of the classical MD.  This shift becomes
more pronounced with decreasing $n$. Furthermore also the shape of the
distributions starts to become different if $n$ is small.

\begin{figure}[t]
\psfig{file=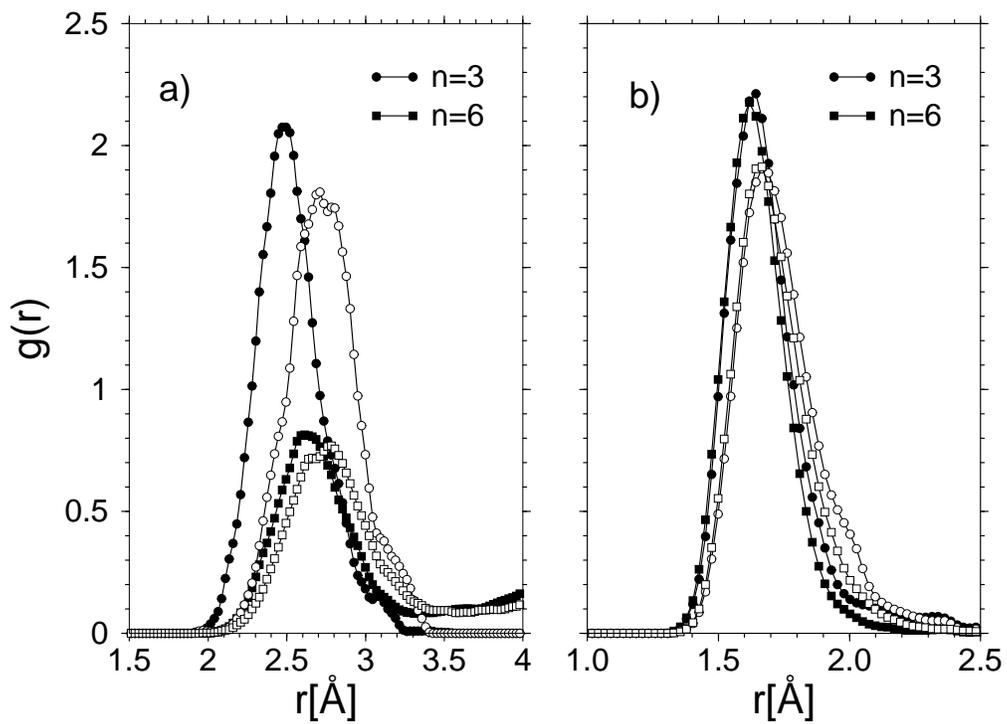,height=9cm}
\vspace{0.3cm}
\caption{\label{fig8}Radial distribution function for different ring sizes,
         (a) O--O and (b) Si--O (BKS: filled symbols, CPMD: open symbols).}
\end{figure}

This effect can be understood better if one analyzes the partial radial
distribution functions $g(r)$ which are shown in Fig.~\ref{fig8}. We see
that for the Si--O correlation the curves from the CPMD are shifted to larger
distances by about $0.04$~\AA~and that this shift is independent of
$n$. Also the $g(r)$ for the O--O correlation are shifted to larger $r$, but
now we observe different shifts for different $n$. In particular we note that
the O--O distance is nearly independent of $n$ for the CPMD whereas it
increases with $n$ in the case of the BKS result. These effects result in
the difference of the distribution of the O--Si--O angles if $n$ is small.

In conclusion we find that for the structure on larger length scales the
BKS simulation and the CPMD yield similar results whereas the details
of the structural elements on short scales (for instance, the short rings)
are different in both methods. This shows that it is probably necessary to
use {\it ab initio} methods like CPMD if one wants to reproduce quantitatively the
properties of silica surfaces on short length scales.
\vspace*{-0.2cm}

\section{Interfaces between liquid and crystalline silica}

This section is devoted to a classical MD of an interface between liquid
and crystalline silica where we have also used the BKS potential to model
the interactions between the Si and O atoms. A straightforward method
to produce the latter interface would be to wait for the formation of
a crystalline nucleus in a supercooled melt. But presently this is an impossible
task since the viscosity of a silica melt at the melting temperature,
$T_{\rm m}\approx 2000$~K, is around 10$^7$~Poise and thus the time
scale that can be covered by a MD simulation is by far too short to
observe the growth of a nucleus in a supercooled silica melt.

\begin{figure}[t]
\hspace*{-1.3cm}
\psfig{file=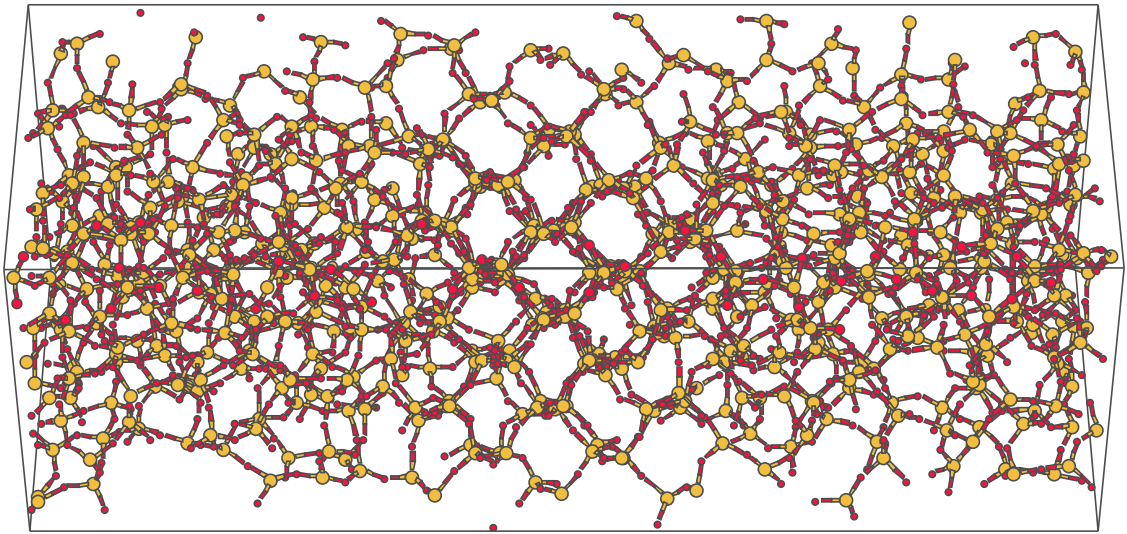,height=8cm}
\caption{\label{fig1tor}Snapshot of the system at $T=3100$~K.}
\end{figure}

Our strategy to prepare a liquid--crystalline interface is as
follows: In a first step a pure melt and a pure crystal ($\beta$
cristobalite), both containing 1944 particles, are relaxed at $T=3100$~K
for 1.6~ns and 160~ps, respectively. Thereby, the box lengths are
$L_x=L_y=21.375$~\AA~in $x$ and $y$ direction and $L_z=61.465$~\AA~for
the melt and $L_z=64.125$~\AA~for the crystal in $z$ direction. The
corresponding densities are $2.32$~g/cm$^3$ in the case of the melt
and $2.22$~g/cm$^3$ in the case of the crystal.  In a second step 648
particles are removed from the equilibrated liquid configuration such
that a free space is created in the middle of the simulation box. Then
one cuts out a part of the crystal consisting of 648 particles that fits
exactly in the latter free space (see Ref.~\cite{stuehn} for the details)
and one combines the liquid and crystalline pieces. This configuration
is relaxed for about 30~ps and as a result one obtains a system with
liquid and crystalline SiO$_2$ phases that are joined to each other
via two interfaces. A snapshot of such a configuration is shown in
Fig.~\ref{fig1tor}.

\begin{figure}[t]
\hspace*{-1.3cm}
\psfig{file=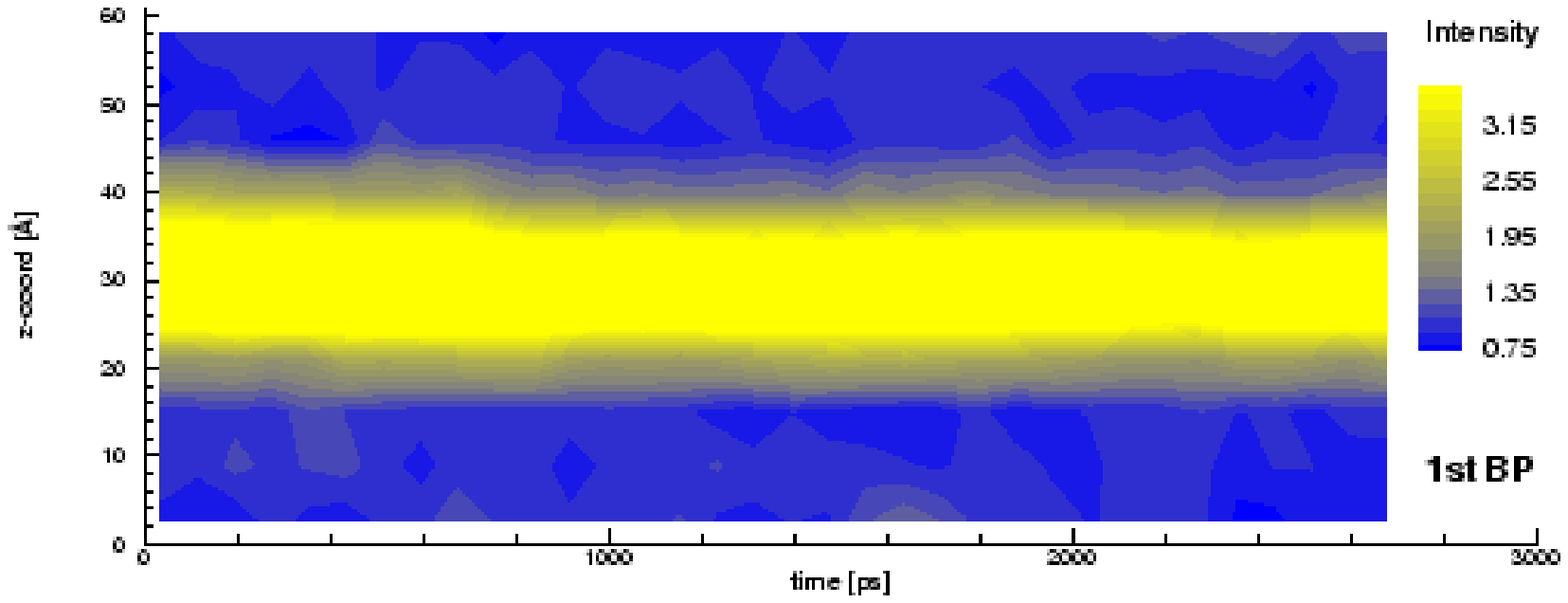,height=6cm}
\hspace*{-1.3cm}
\psfig{file=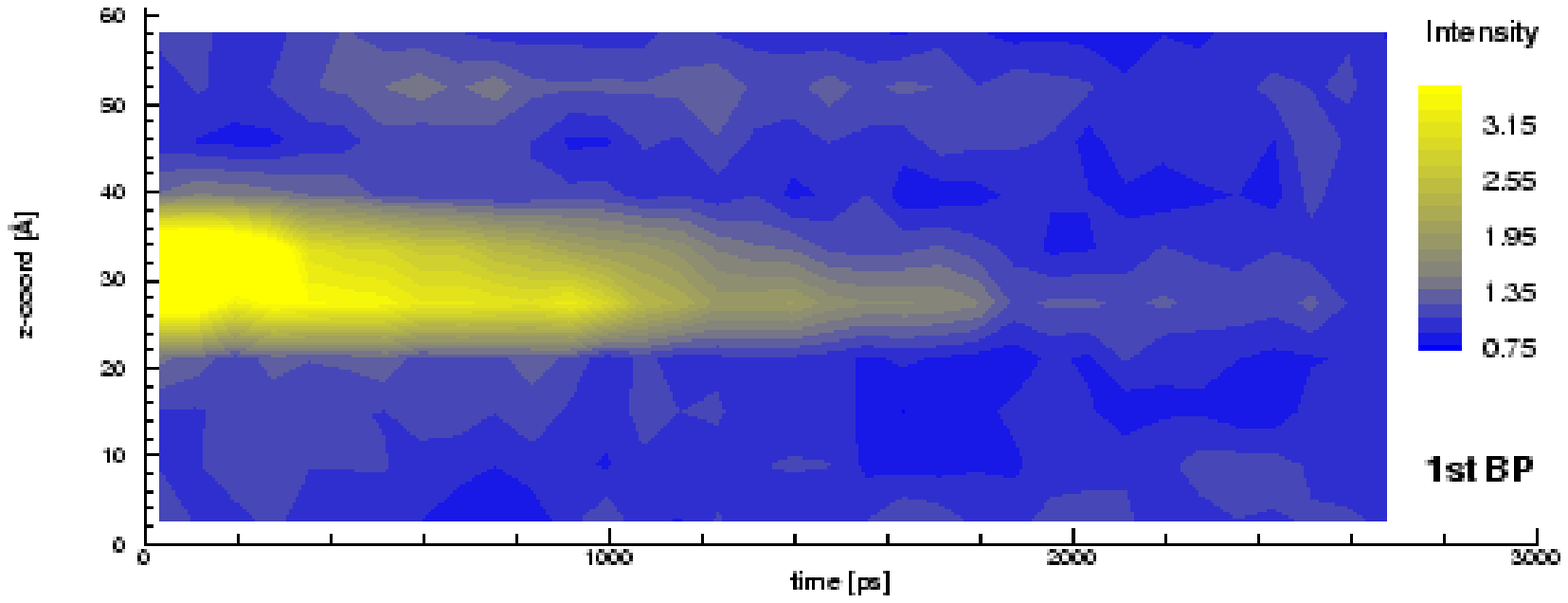,height=6cm}
\caption{\label{fig2tor}Time evolution of the intensity of the
         first Bragg peak in the static structure factor for two different
         samples at the temperature $T=3100$~K.}
\end{figure}

After the preparation procedure a microcanonical run is started
where one expects that the crystal in the middle of the system melts
because the temperature $T=3100$~K is significantly above the melting
temperature of our system~\cite{roder01}. A good order parameter to
quantify the melting of the crystal is the intensity of the first Bragg
peak in the static structure factor~\cite{stuehn}. The latter is shown
in Fig.~\ref{fig2tor} for two different samples. Whereas in the first
sample (upper plot of Fig.~\ref{fig2tor}) the crystal does not melt at
all even after about 2.6~ns, in the same time span the second sample the
crystal is completely melted. More details on that can be found in 
Ref.~\cite{stuehn}.
\vspace*{-0.2cm}

\section{A silica melt between walls}

We turn now our attention to a silica melt between walls to study the
wall--fluid interface structure and to see how the melt behaves
under shear. As a model to describe the interactions between the atoms in
silica we use again the BKS potential. The walls were not constructed as
to model a particular material but rather a generic surface that can be
simulated conveniently. Each wall consisted of 563 point particles forming
a rigid face--centered cubic lattice with a nearest--neighbor distance
of $2.33$~\AA. These point particles interact with the atoms in the fluid
according to a 12--10 potential, $v(r)= 4 \epsilon \left[ (\sigma /r)^{12} -
(\sigma /r)^{10} \right]$ with $\sigma=2.1$~\AA, $\epsilon=1.25$~eV, $r$
being the distance between a wall particle and a Si or O atom. The main
details of the MD simulations are as follows: The simulation box had
linear dimensions $L_x=L_y=23.066$~\AA~in the directions parallel to
the walls (in which also periodic boundary conditions were applied),
and $L_z=31.5$~\AA~in the direction perpendicular to the walls.
Thus, $N=1152$ atoms (384 Si atoms and 768 O atoms) were contained
in the system to maintain a density around $2.3$~g/cm$^3$ which is
close to the experimental one at ambient pressure. All runs were
done in the NVT ensemble whereby the temperature was kept constant by
coupling the fluid to a Nos\'e--Hoover thermostat~\cite{binder03}. We
investigated the temperatures $T=5200$~K, $T=4300$~K, and $T=3760$~K
at which we first fully equilibrated the system for 29, 65, and 122~ps,
respectively. At $T=5200$~K and $T=3760$~K we continued with additional
runs over 164 and 490~ps, respectively, from which we analyzed the
equilibrium structure. Then we switched on a ``gravitational'' field of
strength $a_e=9.6$~\AA/ps$^2$ that was coupled to the mass of the
particles.  With the acceleration field, runs were made over 736~ps,
1.23~ns, and 3.27~ns at $T=5200$~K, 4300~K and 3760~K, respectively. In
addition at $T=5200$~K we did a run over 1.72~ns with field strength
$a_e=3.8$~\AA/ps$^2$. The total amount of computer time spent for these
simulations was 16 years of single processor time on a Cray T3E. More
details on the simulation can be found in Ref.~\cite{horbach02}.

\begin{figure}[t]
\psfig{file=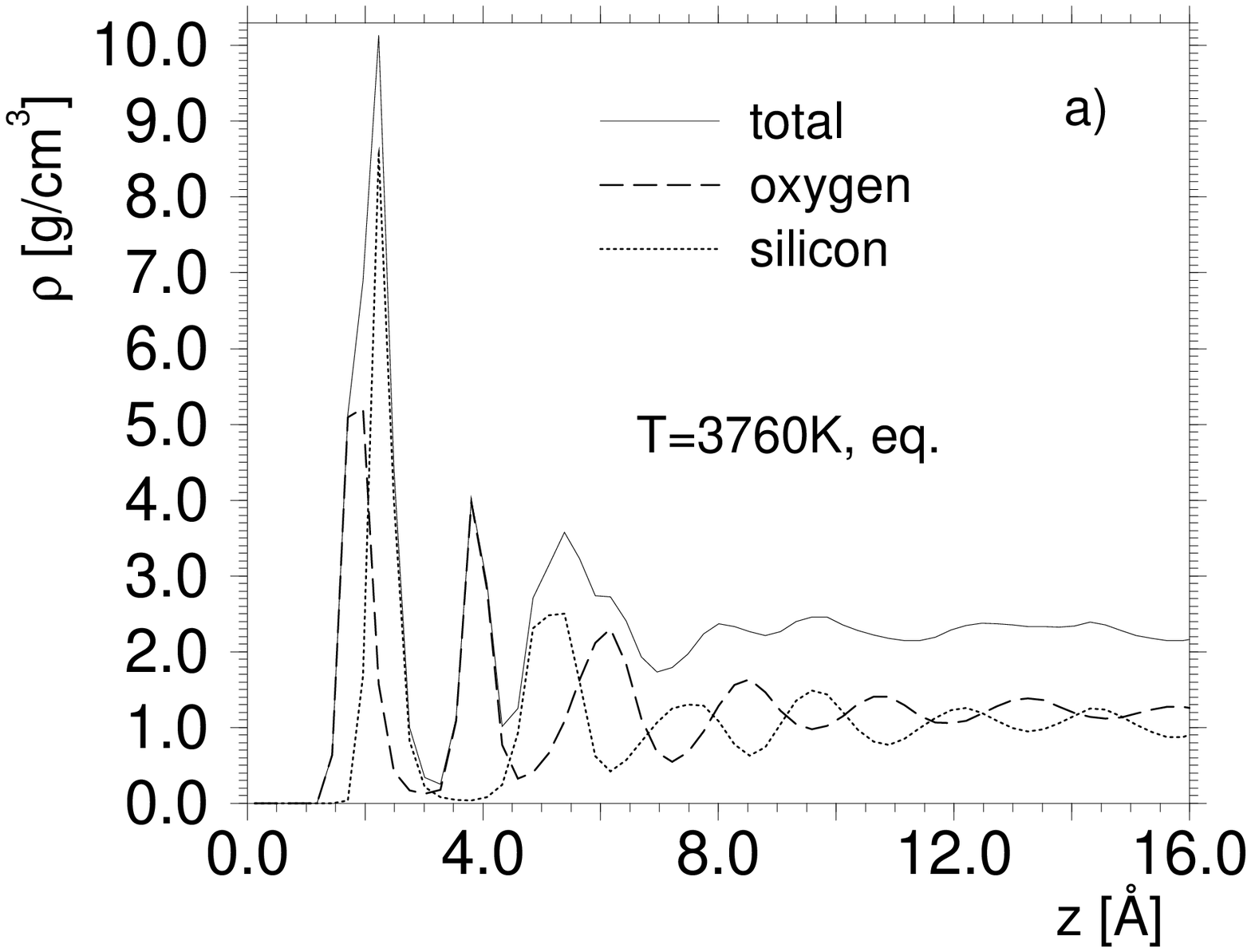,height=8cm}
\psfig{file=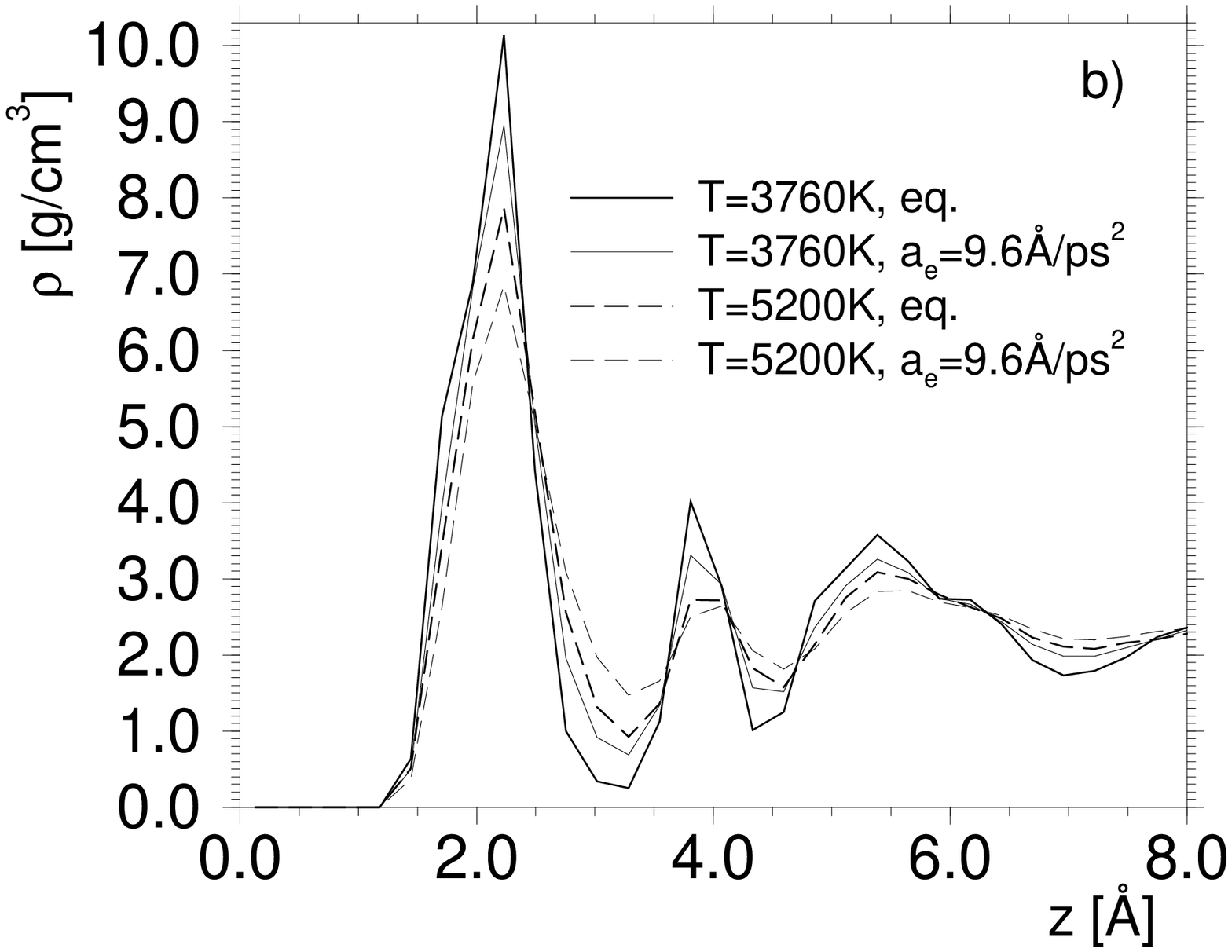,height=8cm}
\caption{\label{fig2jcp}(a) Total density profile and partial density
         profiles for oxygen and silicon in a system at equilibrium
         at $T=3760$~K. (b) Total density
         profiles at equilibrium and with acceleration field $a_e=9.6$~\AA/ps$^2$
         for the indicated temperatures.}
\end{figure}

Fig.~\ref{fig2jcp}a shows the density profiles across half of the film
for all atoms and for the oxygen and silicon atoms only. In contrast to
typical density profiles in simple monoatomic liquids, the oscillations of
the total profile, which indicate a layering near the walls, do not have a
regular character. This behavior can be explained by the partial density
profiles for oxygen and silicon.  Obviously, the tetrahedra adjacent
to the walls prefer to be aligned such that a two--dimensional plane forms which
contains three out of the four oxygen atoms of a tetrahedron, as well as
the silicon atom at their center (slightly further away from the wall),
while the fourth oxygen atom of the tetrahedron has to be further away
from the wall for geometrical reasons: this causes the second peak of the
oxygen distribution.  Thus, the walls have a tendency to ``orient'' the
network of coupled SiO$_4$ tetrahedra in the fluid locally. It is evident
that the oscillations in the local density of both silicon and oxygen
are rather regular, like a damped cosine function, but the wavelength
and phase of both cosine functions are different: their superposition
causes then the rather irregular layering structure of the SiO$_2$
total density. We expect that similar effects also occur in many other
associating molecular fluids confined between walls if the wall--fluid
interaction is weak enough that it does not affect the chemical ordering
in the fluid as it is the case in our system.

For $z > 8$~\AA~the total density profile shows only small oscillations
around a constant value of 2.3~g$/$cm$^3$ which is an indication for bulk
behavior. Thus, keeping in mind that the density profile is symmetric,
the bulk in our system seems to extend in $z$ direction from about
$8$~\AA~to $23.5$~\AA, i.e.~it has a width of about 15.5~\AA. 

The behavior of the total density profiles at the temperatures $T=5200$~K
and $T=3760$~K in equilibrium and with an external force with an
acceleration of $9.6$~\AA$/$ps$^2$ can be seen in Fig.~\ref{fig2jcp}b. In
equilibrium the effect of decreasing temperature on the oscillations near
the wall is an increase of the peak heights that is accompanied with a
smaller value of $\rho(z)$ at the minima between the peaks. Thus, the
layering becomes more pronounced if one decreases the temperature. If
one switches on the gravitational--like field the effect is similar to an
increase of the temperature. In the bulk region the density profiles are
not very sensitive to a variation of temperature and/or the presence
of the external force. Within the accuracy of our data the same
value of about $2.3$~g$/$cm$^3$ is reached for all four cases under
consideration. 

\begin{figure}[t]
\hspace*{-0.5cm}
\psfig{file=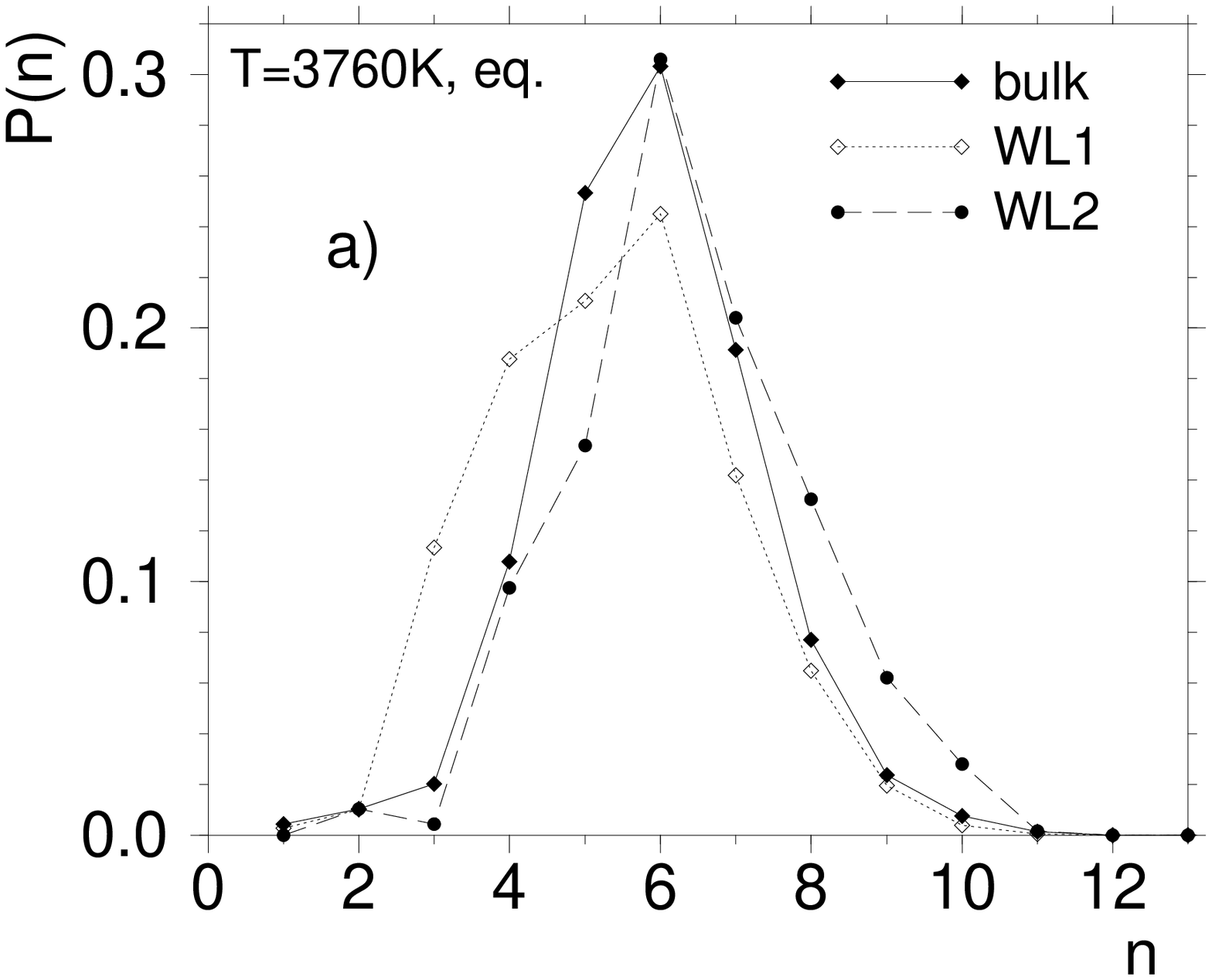,height=5cm}
\hspace*{0.5cm}
\psfig{file=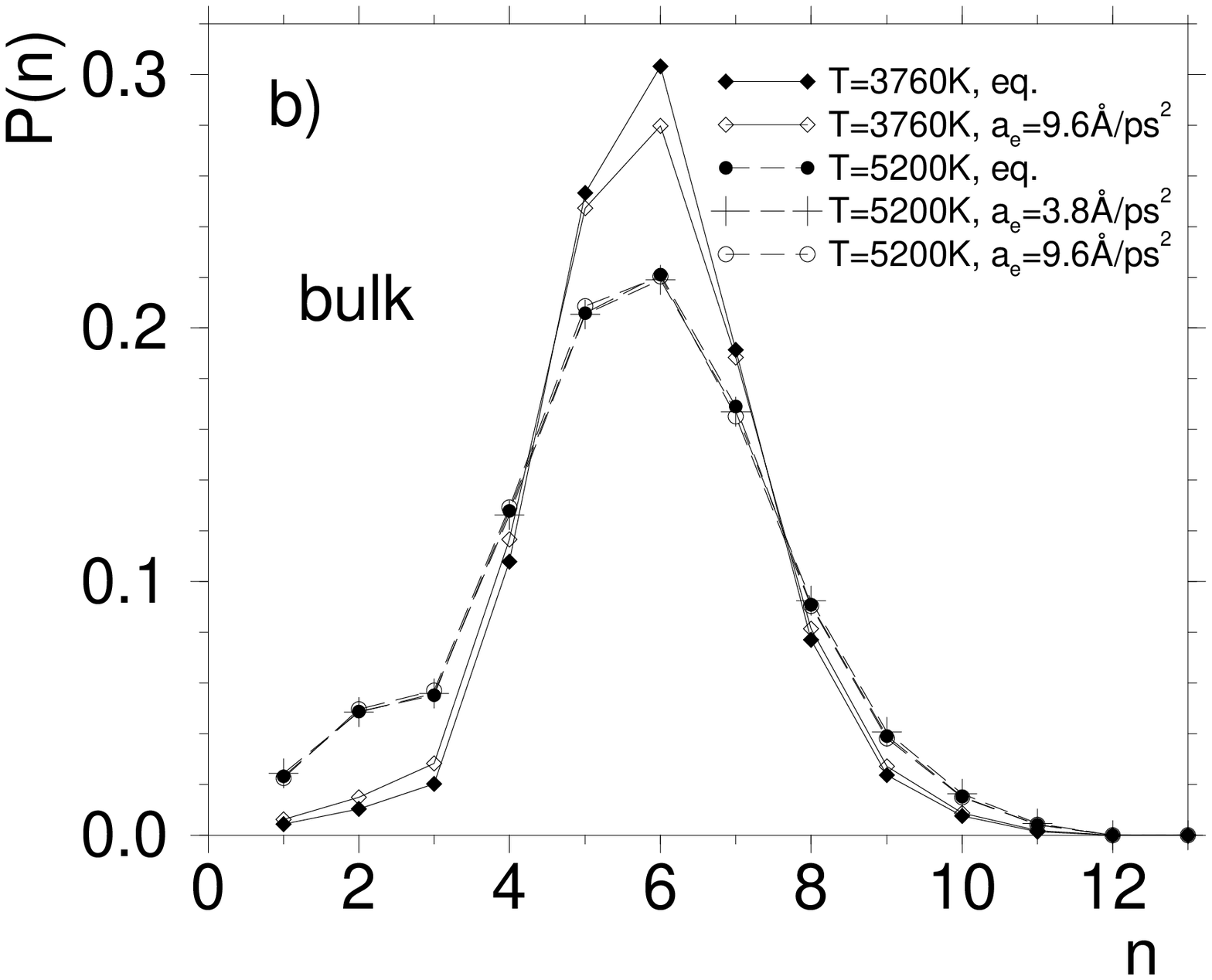,height=5cm}
\hspace*{-0.5cm}
\psfig{file=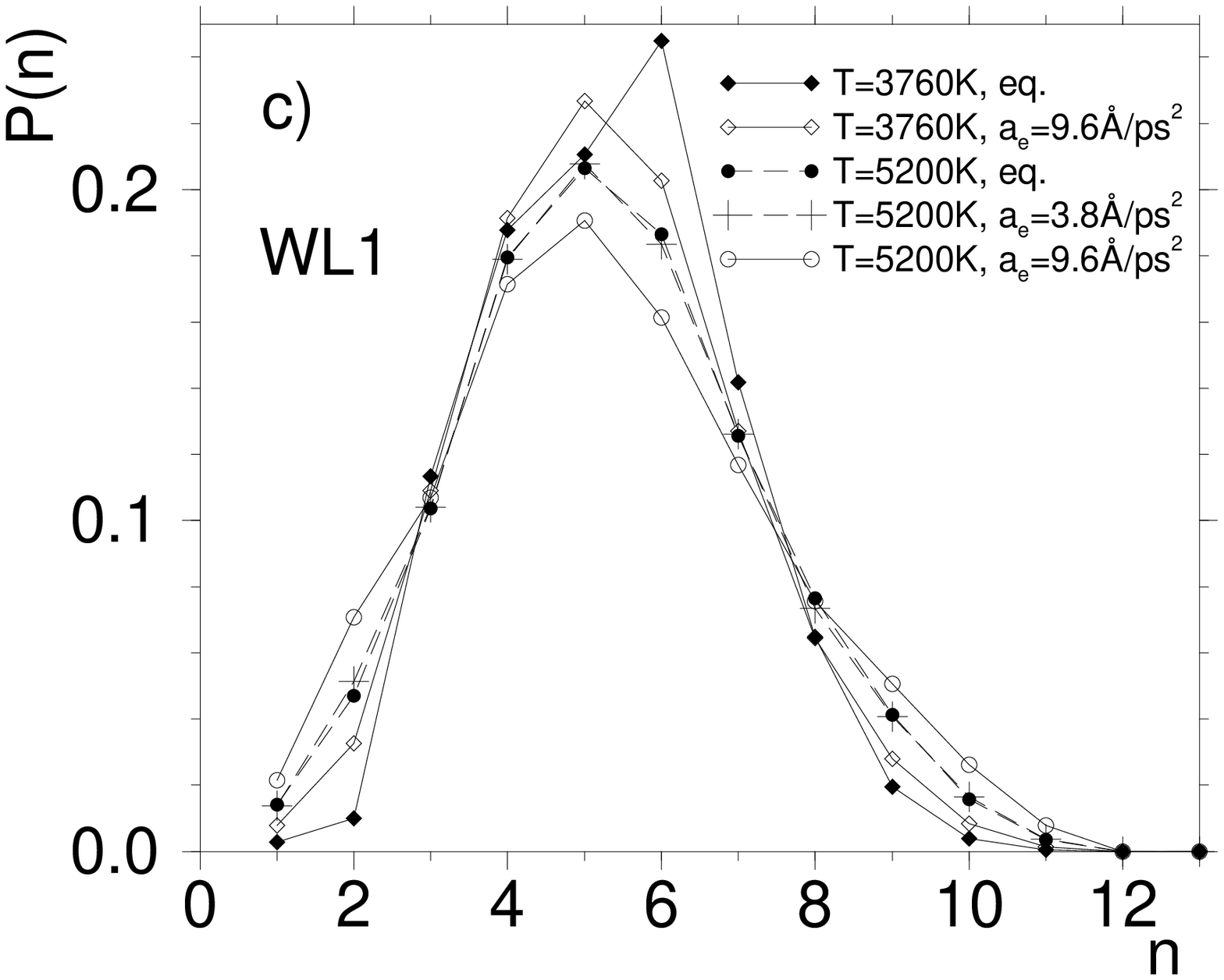,height=5cm}
\hspace*{0.5cm}
\psfig{file=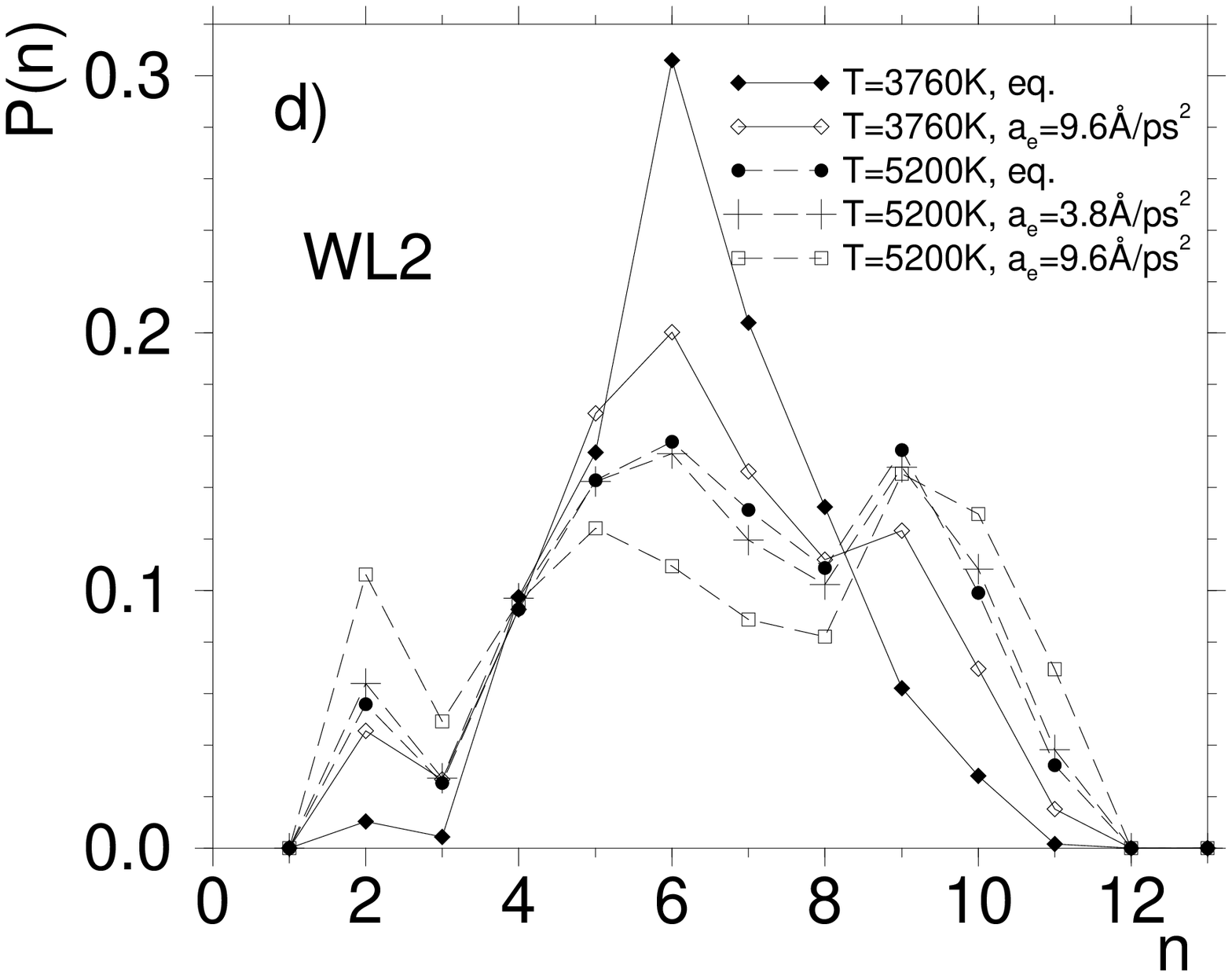,height=5cm}
\vspace*{0.2cm}
\caption{\label{fig5jcp}Probability $P(n)$ that a ring has a length $n$,
         a) at $T=3760$~K in equilibrium for the bulk, WL1, and WL2,
         b) at $T=3760$~K and $T=5200$~K in equilibrium
         and under the indicated acceleration fields in the bulk,
         c) the same as in b) but for WL1, and
         d) also the same as in b) but for WL2. See text
         for the definition of WL1 and WL2.}
\end{figure}

Fig.~\ref{fig5jcp}a shows the ring distribution function $P(n)$
for $T=3760$~K in the bulk and in two different wall layers denoted
by WL1 and WL2. WL1 and WL2 are defined as the regions which are
respectively within a distance of $6.25$~\AA~and $3.0$~\AA~away from
the wall corresponding to the second minimum of the density profile for
silicon and the first minimum of the total density profile, respectively
(see Fig.~\ref{fig2jcp}a). In each region, i.e.~bulk, WL1, and WL2, we
took only those rings into account that fit completely into it. Thus,
in WL2 those rings are counted that are formed at each case by the first
and the second O and Si layers (with respect to the distance from the
wall), whereas with WL2 only those rings are taken into account that
are formed by the first O and the first Si layer.  This is justified
because the first two oxygen and the first two silicon layers are
well--defined in that the minima in the corresponding density profiles
are close to zero density in the case of WL2 and around the small value
$\rho=0.5$~g$/$cm$^{3}$ in the case of WL1. Furthermore, we can infer
from Fig.~\ref{fig2jcp}a that in contrast to the second layers the
first oxygen layer overlaps strongly with the first silicon layer and
the overall thickness of both layers is only about $2$~\AA.  Thus, the
first oxygen and the first silicon layer form a quasi--two--dimensional
plane and $P(n)$ for WL2 gives a distribution of rings that have an
orientation parallel to the walls whereas in $P(n)$ for WL1 also the
rings perpendicular to the walls are included.

In bulk simulations of SiO$_2$ one finds a maximum around 
$n=6$~\cite{vollmayr96}. This is plausible since in silica the high--temperature
crystalline phase at zero pressure, $\beta$--cristobalite, exhibits only
six--membered rings.  In WL1 the probability for $n\ge 6$ is smaller
than in the bulk in favor of a relatively high probability of $n=3$
and $n=4$. In WL2 it is the other way round: $n=4$ and even more $n=5$
are less frequent than in the bulk in favor of $n=8,9,10$. The ring
structure near the walls that corresponds to these findings is as follows:
Perpendicular to the walls small rings with $n=3,4$ are seen such that,
e.g., $n=3$ is formed by two silicon atoms from the first silicon layer
with a third one from the second silicon layer. In contrast to that,
parallel to the walls (considering the first oxygen and the first silicon
layer) an open structure with relatively large rings is observed which
compensates somewhat the dense packing of SiO$_4$ tetrahedra perpendicular
to the walls.

Fig.~\ref{fig5jcp}b shows the behavior of $P(n)$ in the bulk at the
two temperatures $T=3760$~K and $T=5200$~K in equilibrium and under
shear. We can immediately infer from the figure that the considered
shear fields have only a small effect on the structure. At $T=5200$~K one
has a relatively large amount of two-- and three--membered rings and their
frequency is more than a factor of two smaller at $T=3760$~K. In
Ref.~\cite{vollmayr96} it was shown that their frequency of occurrence
decreases further with decreasing temperature such that the amount
of two--membered rings falls far below 1\% for systems that have typical
structural relaxation times of the order of $1$~ns.

In contrast to the bulk in WL1 significant changes in the
ring distribution take place if the system is sheared (see
Fig.~\ref{fig5jcp}c), and the external force field affects the ring
structure such that small and large rings are formed, while at the
same time especially the amount of six--membered rings decreases. Only
for the smaller field strength $a_{{\rm e}}=3.8$~\AA$/$ps$^2$ there
are no significant changes in the ring distribution. As we have shown
elsewhere~\cite{horbach02} the latter is accompanied with a very small
slip motion at the walls whereas a large slip velocity is correlated
with strong rearrangements in the ring structure.  One can also infer
the remarkable fact from Fig.~\ref{fig5jcp}c that the probability to find
rings with $n=3,4$ does not change very much when an external acceleration
field is switched on. This is reasonable because these small--membered
rings, as we have seen before, are located perpendicular to the walls
and thus they are very stable to shear forces that are imposed parallel
to the walls.

The strongest rearrangements in the ring structure due to a shear field
are found when we consider the region WL2. The corresponding curves are
shown in Fig.~\ref{fig5jcp}d. Again, there are only minor changes in $P(n)$
at $T=5200$~K and $a_{{\rm e}}=3.8$~\AA$/$ps$^2$ as compared to the
corresponding equilibrium case which is, as mentioned before, related
to the presence of only a very small slip velocity. For the higher
acceleration field, $a_{{\rm e}}=9.6$~\AA$/$ps$^2$, the ring structure
becomes more heterogeneous and the effect of the external field is if one
would locally increase the temperature. The rearrangements in the ring
structure can be summarized as follows: Rings mainly of size $n=6,7,8$ are
broken under the influence of the shear force and instead small rings with
$n<4$ and very large rings with $n\ge 9$ are formed.  
\vspace*{-0.2cm}

\section{Summary}
We have shown that large scale MD simulations are able to give a lot of
insight into the microscopic surface and interface structure of silica. In
all three examples that we have presented in this report it is necessary
to simulate relatively large systems on a relatively large time scale and
thus the use of parallel supercomputers is indispensable. In our study
of the free silica surface the Car--Parrinello MD was used for which
length and time scales that can be covered are even very restricted on a
parallel computer (CPMD simulations are typically on a ps time scale for
systems of about 100 particles). Thus, the development and application
of more powerful parallel computers is required to gain further insight
from atomistic simulations.

Acknowledgments:
We thank the HLRZ Stuttgart for a generous grant of computer time on
the CRAY T3E. C. M. and T. S. are grateful to SCHOTT Glas for partial
financial support.

\clearpage


\end{document}